# Quantifying COVID-19 enforced global changes in atmospheric pollutants using cloud computing based remote sensing

Manmeet Singh[1,2], Bhupendra Bahadur Singh[1,3], Raunaq Singh[4], Badimela Upendra[5], Rupinder Kaur[6], Sukhpal Singh Gill[7], Mriganka Sekhar Biswas*[1,8]

manmeet.cat@tropmet.res.in, bhupendra.cat@tropmet.res.in, raunaqsingh126@gmail.com,

upendra.b@ncess.gov.in, rupinderchem@gmail.com, s.s.gill@qmul.ac.uk,

mriganka.cat@tropmet.res.in

[1] Centre for Climate Change Research, Indian Institute of Tropical Meteorology, Pune, India, Ministry of Earth Sciences, Government of India

[2] IDP in Climate Studies, Indian Institute of Technology, Bombay, India

[3] Department of Geophysics, Banaras Hindu University, Varanasi, India

[4] School of Sciences, Indira Gandhi National Open University, Delhi, India

[5] National Centre for Earth Science Studies, Thiruvananthapuram, India, Ministry of Earth Sciences, Government of India

[6] Department of Chemistry, Guru Nanak Dev University, Amritsar, India

[7] School of Electronic Engineering and Computer Science, Queen Mary University of London, London, United Kingdom

[8] Department of Atmospheric and Space Sciences, Savitribai Phule Pune University, Pune, India

*Corresponding Author

Mriganka Sekhar Biswas
Email: mriganka.cat@tropmet.res.in
Centre for Climate Change Research
Indian Institute of Tropical Meteorology
Pashan, Pune 411008, India

**Abstract**

Global lockdowns in response to the COVID-19 pandemic have led to changes in the anthropogenic activities resulting in perceivable air quality improvements. Although several recent studies have analyzed these changes over different regions of the globe, these analyses have been constrained due to the usage of station based data which is mostly limited upto the metropolitan cities. Also the quantifiable changes have been reported only for the developed and developing regions leaving the poor economies (e.g. Africa) due to the shortage of in-situ data. Using a comprehensive set of high spatiotemporal resolution satellites and merged products of air pollutants, we analyze the air quality across the globe and quantify the improvement resulting from the suppressed anthropogenic activity during the lockdowns. In particular, we focus on megacities, capitals and cities with high standards of living to make the quantitative assessment. Our results offer valuable insights into the spatial distribution of changes in the air pollutants due to COVID-19 enforced lockdowns. Statistically significant reductions are observed over megacities with mean reduction by 19.74%, 7.38% and 49.9% in nitrogen dioxide ($NO_2$), aerosol optical depth (AOD) and $PM_{2.5}$ concentrations. Google Earth Engine empowered cloud computing based remote sensing is used and the results provide a testbed for climate sensitivity experiments and validation of chemistry-climate models. Additionally, Google Earth Engine based apps have been developed to visualize the changes in a real-time fashion.

## 1. Introduction

The impedance caused by the COVID-19 pandemic has led to worldwide disruptions in day-to-day human activities across the globe. As per the World Health Organization (WHO) Weekly Epidemiological Update issued on 15th December 2020, more than 70 million were infected alongside death numbers exceeding 1.6 million worldwide. The characteristics of the virus included rapid spread from human-to-human making its infections difficult to contain. There has been an evidence of virus spreading through the air, after the WHO declared it a pandemic on 30th January 2020 (Zander et al, 2020). Almost every country was affected with frequent cases of infected but asymptomatic individual. Those potential virus carriers made its transmission hard to track. In that scenario and unavailability of any vaccine hitherto, most of the countries declared lockdowns to prevent the spread of novel coronavirus. This led to a halt in the anthropogenic activities in urban and industrialized areas across the world.

Estimates from the United Nations show that 55% of the world's population lives in urban areas as of 2018, and by 2050 close to 68% would be living in urban areas (The Department of Economic and Social Affairs, 2019). By 2018, there are 33 megacities in the world with population more than 10 million, hosting ~12.5% of the world's total urban population. Due to the high population density, understanding the environmental impacts of megacities are of utmost concern for policy makers to ensure public health and safety. Emissions from the industries, automobiles, construction processes, and other anthropogenic activities have led to high levels of air pollution over megacities (Butler et al., 2008; Gurjar et al., 2008; Molina et al., 2012; Baklanov et al., 2016; Marlier et al., 2016), making them hotspot for various greenhouse gases, ozone precursors and aerosols observed well upto the tropopause layers (Brunamonti et al., 2018; Hanumanthu et al., 2020). Various studies have reported increased mortality rates around the world's megacities caused by the air pollution. Cohen et al. (2017) reported that exposure to $PM_{2.5}$ and tropospheric ozone ($O_3$) caused ~4·2 and ~0.25 million deaths worldwide respectively in 2015. Gaseous pollutants like nitrogen dioxide ($NO_2$), sulfur dioxide ($SO_2$), and carbon monoxide (CO) are also responsible for human health hazards (WHO, 2013; US Environmental Protection Agency, 2015). The primary sources of nitrogen dioxide in the atmosphere are fossil fuel and biomass burning and various microbiological reactions in wildfires, lightning and soils. In addition to the anthropogenic sources, natural emissions such as that from the volcanic eruptions are also harmful to the environment in the short term and also modulating the global hydroclimate (Singh et al. 2020).

Nitric oxide (NO) rapidly oxidizes to $NO_2$ in the air and plays crucial role in the formation of photochemical smog, affects air quality and climate. The dominant impact of NOx ($NO+NO_2$) emissions on the climate is through the formation of $O_3$, the third largest single contributor to positive radiative forcing. Emissions of $NO_x$ generate indirect negative radiative forcing by shortening the atmospheric lifetime of $CH_4$. $NO_x$ dominantly controls the $O_3$ budget through photochemistry. It is well documented that high amounts of NOx emitted from the industries, thermal power plants and automobiles leads to surface ozone production (Lelieveld et al., 2000) and leads to the exceedances of the standard limit of the urban ozone concentration (Frost et al., 2006). Tropospheric NO and $O_3$ are potential greenhouse gases and influence the lifetime of the other greenhouse gases. Also, the $NO_x$ compounds act as precursors for the aerosol nitrate and influences significantly the abundance of the hydroxyl radical (OH). Furthermore, $NO_2$ also produces nitric acid ($HNO_3$) which is a major component of acid rain by reacting with hydroxyl radical (OH). The sources of $NO_x$ include both natural and anthropogenic sources. The natural sources include atmospheric flux exchanges, lighting activity, soil emissions, and forest and grassland fires while the anthropogenic sources are primarily through emissions from the power plants, transportation (automobiles, ships and aircrafts), industrial emissions and biomass burning (Guha et al., 2020). The lifetime of NOx is of the order of minutes to hours, which depends on various factors such as the season, location, photolysis rate and the concentration of hydroxyl radical (Lamsal et al., 2010). Hence, the short life time and the inhomogeneous source distribution of $NO_x$ clearly lead to the spatiotemporal variations in the $NO_2$ concentration in the troposphere and it is well proven that tropospheric $NO_2$ observed from the space is dominated by the amount in the boundary layer (Ghude et al., 2008). Total global $NO_x$ emissions have increased from an estimated pre-industrial value of 12 TgN yr$^{-1}$ (Holland et al., 1999; Galloway et al., 2004) to between 42 and 47 TgN yr$^{-1}$ in 2000 (Solomon et al., 2007). The range of surface $NO_x$ emissions (excluding lightning and aircraft) used in the current generation of global models is 33 - 45 TgN yr$^{-1}$ with small ranges for individual sources.

Satellite observations in the past few decades have helped to study long-term spatial and temporal variation of pollutants around the world. Hilboll et al. (2013) and Georgoulias et al. (2019) studied the long-term trend of $NO_2$ over the world's megacities using multiple satellite observations. Long term trends of $SO_2$ have also been studied across the world where the Asian countries show mostly increasing trends whereas opposite trends have been reported over North America in the recent decades (Lu et al., 2011, 2013; Kharol et al., 2017; van der A et al., 2017). Using OMI/MLS satellite observation Cooper et al. (2020) reported 8% relative increase in tropospheric ozone burden over northern hemisphere (NH) compared to southern hemisphere (SH) in recent times. The analysis of these satellite datasets and their pre-processing is cumbersome and error prone due to involved steps. Raw satellite datasets are available in different formats for different satellites and are mostly available as swath-based products in case of missions such as MODIS and image blocks in case of Sentinel. The analysis of these datasets requires them to be mapped to gridded form, necessitating the need of transforming to real coordinates which is a challenging task. Moreover, different satellites-based instruments require different corrections and filters to be applied based on data quality over different regions and instrument-to-instrument differences. The requirement of large disk-space is also a challenge for many researchers, due to unavailability of high-end computational resources at their end.

Cloud computing offers hope in solving these challenges by providing cloud as a service platform for preprocessing, analyzing and visualizing big data (Gill et al., 2019). Google has introduced a new cloud-based platform called Google Earth Engine for efficient and fast processing of large geospatial datasets. It provides a systematic platform to analyze planetary-scale geospatial data to uncover robust computational capacities of Google which can be used as a source for analyzing problems and proposing solutions for environmental protection, climate monitoring, water management, food security, disease, disaster, drought and deforestation. Moreover, this engine is now being utilized to distribute and share results with others, to develop mobile apps or web based services, interpreting various types of geospatial data, and for assessing land use change, monitoring climate, malaria risk mapping, flood mapping, urban mapping, rice paddy mapping, crop yield estimation, global surface water change and global forest change. For example, an application based on Google Earth Engine is developed to detect land cover change, which has been implemented successfully in Singapore (Sidhu et al., 2018). Researchers from Ukraine used Google Earth Engine and created a high resolution crop classification map for a large spatial region (Shelestov et al., 2017). Further, it has been identified that machine learning techniques such as random forest and linear regression are working efficiently for satellite imagery processing and COVID-19 predictions (Tamiminia et al., 2020; Tuli et al., 2020).

Numerous recent studies have reported improvement in air quality around the world due to large scale lockdown as a preventive measure to contain COVID-19. Muhammad et al. (2020) reported such an improvement using satellite data, however only over a short span of time i.e., the initial lockdown period. He et al. (2020) studied the short-term impacts of lockdowns across cities in China by analyzing air quality parameters such as $PM_{2.5}$, $PM_{10}$, $NO_2$, $O_3$, CO and $SO_2$ using station based data. Their findings showed that large and wealthier cities in China had a greater reduction in air pollution than otherwise. Huang et al. (2020) showed that impact of COVID-19 forced lockdowns in improving air quality was not always apparent. Shen et al. (2020) studied the changes in air pollution with respect to meteorology by comparing the 2020 lockdown period with the 21-year long term means. They noted that large-scale transport of the pollutants reflected importance of meteorology on air quality at regional scales. Similarly, Chang et al. (2020) attributed enhanced haze creation during lockdown in China to pollutant pathways. Rodríguez-Urrego and Rodríguez-Urrego (2020) reported reduction in $PM_{2.5}$ over 50 most polluted capitals around the world. They found a 12% reduction in $PM_{2.5}$ around the world's most polluted capitals with Bogotá, Colombia showing the highest decrease (57%). Sharma et al. (2020) reported reduction in air pollution in 22 Indian cities during COVID-19 forced lockdown. Similarly, improvement in air quality over central China during lockdown was also reported by Xu et al. (2020). Berman and Ebisu (2020) reported a reduction in $NO_2$ and $PM_{2.5}$ predominantly over the urban United States. They compared early and late/no business closure scenarios in different counties of the country and found clear reductions in the pollutant levels in early closure scenario over the urban counties. Li et al. (2020) used the WRF-CAMX model to assess lockdown induced changes in air quality over Yangtze river delta. They found that though the daily $PM_{2.5}$ reduced during the lockdown, it was still high and more stringent measures were required for better air quality. Menut et al. (2020) also employed WRF-CHIMERE modelling system for understanding the pandemic enforced lockdown changes in air quality over the western Europe during March 2020. Using satellite observations Biswas and Ayantika (2020) reported decrease in $NO_2$, formaldehyde (HCHO), $SO_2$ and Aerosol Optical Depth (AOD) over India during pre-monsoon period (March, April and May) compared to past three years due to COVID-19 induced lockdown. The study reported that mitigation of ozone and large reductions in $NO_2$ were associated with the muted decrease in particulate matter concentrations. However, our survey finds some limitations in these studies. Firstly, most of them have used station data to assess the changes in local and regional air quality which limits the spatial coherence and continuity, particularly the usage of $PM_{2.5}$ has been limited in most of the recent studies. Also, there has been an absence of discussions about the AOD, which provides information about the amount of direct sunlight blocked by various particles/gases when it reaches the surface of Earth. Secondly, the previous analyses have reported the cases from brief lockdowns during the early phase of pandemic thereby limiting the robustness of the conclusion. Thirdly, the published results have mostly discussed the air quality changes in developed and developing city/regions while the relatively poorer regions (e.g. African continental regions) have mostly been ignored. In addition to these factors, the existing works are limited by the spread of the observational networks out of the country capitals or even limited observations in the major cities. For example, the existing station data network of Central Pollution Control Board which has been used for the past studies over India, has a good coverage only over the national capital region around Delhi.

Given its global presence, almost every country in the world has used lockdowns as a preventive measure to contain the further spread of pandemic. As seen in the literature survey, qualitatively, it has been felt that the lockdowns have improved air quality due to a reduction in emissions arising out of suppressed anthropogenic activities. Various studies have used station based datasets to bring out this change whereas others have used satellite datasets to show the improvements in air quality over a limited period of the lockdown. Moreover, the studies have been regional in nature and have also not taken into account the high-resolution panoply of datasets into consideration. Having noted the limitations in the recent studies, the present study utilizes the Google Earth Engine's capabilities to quantify the changes in atmospheric pollutants such as $NO_2$, AOD, $O_3$ and $PM_{2.5}$ during the COVID forced lockdowns globally. We use multi-satellite high resolution datasets spanned over the complete lockdown period until 31 May 2020 to show its environmental affects over 8 major continental parts of the world. Cloud computing based remote sensing via Google Earth Engine is used and lesser reported regions in other studies such as Africa have been included and thoroughly explored. Additionally, we also provide

Google Earth Engine based apps to visualize the changes in air quality over city-scale spatial resolutions.

## 2. Data and Methodology

The Google Earth Engine, which is the first ever cloud computing based platform enabling processing, analysis and visualization of satellite and other datasets for the planet Earth, is extensively used in this study. Google Earth Engine enables the processing, analyzing and visualization of these large datasets on the cloud. A large number of the datasets being made available after quality control. Hence, there is no need of performing any major preprocessing or storage making the analysis a smooth process. This study employs Google Earth Engine extensively for the analysis and visualization of air quality and other datasets. In this work it has served as a nodal point for targeted analysis of the quantitative environmental assessment of COVID-19 enforced lockdowns. The analysis is enabled by direct API calls to the requisite data and there is no need to download the same. Its use has catapulted the large-data processing which is otherwise very tedious and error prone task using the traditional approaches. The access to the platform is provided by an application process and the APIs are available in javascript and python. In this work, the python API has been used for analysis. All the codes for analysis and visualization are available at https://github.com/manmeet3591/gee_lockdown.

Satellite data products *viz* $NO_2$ and tropospheric $O_3$ are used from TROPOspheric Monitoring Instrument (TROPOMI; Veefkind et al., 2012), an instrument onboard Sentinel-5 precursor (Sentinel-5P) satellite. The AOD is obtained from MODIS (Schaaf et al., 2002) and particulate matter less than 2.5μ ($PM_{2.5}$) are also based on the data assimilated MODIS product. The European Space Agency had launched Sentinel-5P on 13 October 2017 as a dedicated satellite to observe air pollution. The datasets are available in two versions i.e. Offline (OFFL) and Near Real-Time (NRTI). NRTI products are available earlier than the OFFL products, however

OFFL products offer better quality than NRTI and hence are used in this work. The datasets from Sentinel-5P and MODIS provide daily measurements if the sky is not cloudy. The data from TROPOMI is accessed using Google Earth Engine and is available from July 2018. Tropospheric ozone concentrations available for the tropical band 20°S-20°N is archived at Google Earth Engine servers using the raw data and cloud slicing (csa) and convective cloud differential (ccd) algorithms. AOD data is used from the version 6 of combined MODIS Terra and Aqua product wherein atmospheric corrected data over land at 1 km horizontal resolution is provided by Google Earth Engine. $PM_{2.5}$ is used from the Copernicus Atmosphere Monitoring Service (CAMS) Global Near-Real-Time accessed using Google Earth Engine after 4D-Var data assimilation using datasets from MODIS. 4D-Var is an advanced analytical method to perform data assimilation. Data assimilation involves combining short-range prediction with in-situ measurements to provide the best approximation of Earth system. COVID-19 lockdown start dates are accessed from the news reports and popular articles. Since lockdowns were still on in countries, the present study considers the period upto 31 May 2020 (Table 1).

We calculate the percentage change of various air quality parameters, which is the percentage difference of aggregated means during the two periods. The two periods are defined as the lockdown period for year 2020 (Y20) and the corresponding epoch in the year 2019 (Y19). For instance, to calculate the percentage change in $NO_2$ over South Asia, we average the $NO_2$ maps over Y19 and Y20 and then obtain [(Y19 - Y20)/(Y19)]*100 as the percentage change in $NO_2$ over South Asia in Y19 relative to Y20. We also select 93 urban cities globally by first selecting the megacities (Mage et al., 1996; Gurjar et al., 2010, 2016; Baklanov et al., 2016; Cheng et al., 2016; Marlier et al., 2016), followed by European Union capitals and then the cities with GDP per capita greater than $ 25000. The changes in air quality parameters over these cities are also performed by the methodology above and averaging over the area of the city. The latitude longitude information of the cities is obtained from www.latlong.net. We define the extent of the

cities by the area of the cities from these central points (latitude, longitude) which is also taken as the region over which we compute the pollutant concentration variations. It is worthy to note that for some countries such as Japan and South Korea which did not enforce lockdowns, the analysis for the cities is not done to bring out the exact impact of lockdowns over the air quality. The coordinates, population and GDP data of the global megacities along with the start and end of the lockdown enforcement dates are available on https://rb.gy/t7jzr8. Interactive maps of meteorological fields such as land surface temperature and surface winds along with four air pollutants showing the absolute values of COVID19 lockdown period in 2020 and the corresponding period from 2019 have been prepared as applications of Google Earth Engine. The links for the same can be found from the section 3.9.

## 3. Results and Discussion

In this section we discuss the environmental changes in 8 major regions of the world due to COVID-19 enforced lockdowns. The subsections describe the changes over Africa, Australia and New Zealand, East Asia, South Asia, Europe, North America, South America, Southeast Asia and the global megacities.

### 3.1 Africa

Surface station-based observations records in various African regions show annual mean $NO_2$ and $O_3$ concentrations to be in the range of 0.9-2.4 ppb and 4.0-14.0 ppb respectively (Adon et al. 2010). However, during the lockdown period, the level of these pollutants reduced significantly over the region. The reduction in $NO_2$ concentrations (Fig. 1a) are observed as blobs over urban areas with more than ~30% decrease over large swathes of South Africa, Botswana, Namibia, Angola, Tanzania, Kenya and coastal countries of West Africa. Algeria and Niger show a decrease by ~20% in $NO_2$ concentrations. Over the Arabian Peninsula the decrease is seen only over urban areas such as Riyadh, Dubai, Muscat, Bahrain, Qatar and Israel. The tropospheric ozone product is available only from 20°S to 20°N and a 20-30% decrease can be seen over regions surrounding Tanzania, Zambia, Angola, Kenya and Congo (Fig. 1c). The African regions usually have high AOD values during the local dry seasons (Boiyo et al., 2016) coinciding with the reduced human activity. This period also overlaps with the lockdown periods (March-May) used in the study. However, we observe a decrease in the AOD (Fig. 1b) over South Africa, Botswana and Angola in Southern Africa by ~30-50%, over Nigeria, Ghana Cote d'Ivoire, Sierra Leone and parts of Niger in Eastern Africa by ~30-50%, over Sudan and Egypt by ~20% and parts of north Libya by more than 50% relative to the same period (as that of lockdown period in 2020) in 2019. In the Middle-East, parts of Saudi Arabia, Iraq and Iran show a decrease by ~30%. In addition, parts of Congo, Tanzania and Kenya also show reduction in AOD when compared with the month before lockdown. $PM_{2.5}$ concentrations across African regions have different characteristics wherein West Africa has higher levels partly arising from dust, while the rest of Africa has anthropogenic factors dominating the overall concentrations (Heft-Neal et al. 2018). During the analysis period, reduction in $PM_{2.5}$ levels can be observed over large parts of Namibia, western South Africa, the entire Arabian Peninsula and North Africa to negligible amounts during lockdown (Fig. 1d). A two-thirds decrease can be seen over other parts of Africa. The relative decrease of ~30% in $NO_2$ over Africa is indeed significant. However, when absolute values are analyzed, for example using Google Earth Engine apps, we can observe significant changes only over South Africa and parts of Namibia, Tanzania and Kenya. Over Angola, Namibia and South Africa decrease in $NO_2$ concentration and in $PM_{2.5}$ are almost anti-correlated. We observe that over some regions of Africa, although the change might seem to be large, the base value is not substantial, which can be clearly understood from the Google Earth Engine apps.

### 3.2 Australia and New Zealand

The regions in Australia and New Zealand have strong seasonal cycles in the atmospheric pollutants' levels. The forest fire/burning lead to enhanced levels apart from the contribution from human activities (Reisen et al. 2012). The local autumn season coincided with the analysis period wherein $NO_2$ concentration reductions by ~30% can be seen over urban areas such as Melbourne, Sydney, Auckland, large swathes of New South Wales, Canterbury, Otago and Southland (Fig. 2a). The region does not show much change in AOD except for areas in Brisbane, Manawatu-Wanganui and neighborhoods of Perth and Auckland (Fig. 2b). The tropospheric ozone data is available only for the Northern part of Australia and a reduction by 15-20% can be seen throughout the region (Fig. 2c). $PM_{2.5}$ shows large reductions (upto 100% in many regions) over large parts of Australia and New Zealand. Eastern Australia shows two-third reductions in $PM_{2.5}$ concentrations (Fig. 2d). The decrease in $NO_2$ concentration and $PM_{2.5}$ are almost anti-correlated over Australia and New Zealand. We observe a significant increase in AOD over western and south-western Australia in 2020 which is due to the forest fires in the end of 2019 and the beginning of 2020. However, the signal is not observed in PM2.5 which requires further investigation.

### 3.3 East Asia

Air pollution has increased over east Asia during the last few decades. The rise has come amidst the industrial growth leading to higher emissions mostly from the regions in China and Korea (Kim et al. 2011; Wang et al. 2017). The concentrations of air pollutants have increasing interannual trend and summer season show increased intra-seasonal variations (Jacob and Winner, 2009). Here in our study we note that $NO_2$ concentrations show a high reduction of ~50% in the urban areas of China with the neighborhoods characterizing a reduction of ~33%. South Korea, North Korea and Japan show ~30-40% decrease in $NO_2$ levels (Fig. 3a). It can be seen that AOD reduced by ~50% over Shanghai, parts of South Korea, Beijing and regions around Xi'an. The regions around these centers show ~33% reduction in AOD (Fig. 3b). Tropospheric ozone data is not available (Fig. 3c) from the satellite over this region so we skip that analysis. $PM_{2.5}$ shows a reduction by ~33% over most of the parts of China, North Korea, South Korea and Japan with more than 50% decrease over Mongolia and the surrounding regions (Fig. 3d). It is to be noted that the color scale in Figure 3a-d for relative change is from -50% to +50%. Mongolia shows a significant relative decrease in $NO_2$ which is however not much valuable considering very low base values, and western China show a significant decrease of $PM_{2.5}$. Overall, the lockdown effects in the air pollution are pronounced over east Asia which resulted in improvement of the air quality.

### 3.4 South Asia

Similar to the east Asian region, the air pollution in south Asia has also gone up in the last few decades (Mahajan et al., 2015; Fadnavis et al., 2020 and references therein). The region is characterized by several developing economies that have paced up the industry thereby increasing the air pollutants. The region has strong seasonal variability of air pollutants dominated by the regional meteorological factors (Tiwari et al. 2013; Wang et al. 2017). Particularly the $PM_{2.5}$ concentration minima are noticed in monsoon season which rises during the winter time and is maintained until pre-monsoon. Surface pollutants over the region are dominated by the changes over Indian subcontinent. The COVID-19 spread and subsequent lockdowns in 2020 were imposed during the summer and pre-monsoon months. The changes in $NO_2$ levels show reductions by more than ~30% seen all over Tier I and II cities in India which are concentrated over these urban areas (Fig. 4a). The neighboring countries also show similar characteristics with Lahore, Islamabad, Karachi and Dhaka seeing around ~30% reductions in $NO_2$ concentrations. AOD reductions can only be observed over South India, Indo-Gangetic plains, West Bengal and Myanmar by ~30-40% (Fig. 4b). For the limited region in South India over which Tropospheric ozone data is available, we do not see any change in the concentrations (Fig. 4c). A remarkable change is noticed in $PM_{2.5}$ concentrations where reductions by ~60-70% are noted over large swathes of South Asia with the reductions nearing complete decimation of the species over western Rajasthan in India (Fig. 4d).

## 3.5 Europe

Past studies have shown that the air pollutants have distinct regional variability across Europe. In general $NO_2$, $NO_x$ and $PM_{2.5}$ concentrations are found to be higher in Southern Europe while lesser values are found over the regions in Western and Northern Europe (Eeftens et al. 2012). The region is dominated by the street/urban background concentration ratios for $PM_{2.5}$ along with non-tailpipe emissions. Though meteorology plays an important role in the air pollutants variability over Europe, this region has registered decreasing trends in most of the anthropogenic induced pollutants due to air pollution controls (Barmpadimos et al. 2012; Wang et al. 2017). With specific measures to curb the emissions, long term observations show that AOD has decreased over Europe with largest AOD variations occurring during winter and spring followed by some reductions in the summer and autumn months (Chiacchio et al. 2011). During the analysis period. We notice reductions by ~30% in the major cities of Europe such as Lisbon, Madrid, Barcelona,

Toulouse, Monaco, Manchester, Birmingham, Istanbul, Moscow, Stockholm, Oslo, Helsinki and large parts of Germany, regions in and surrounding Paris and London (Fig. 5a). Lowered AOD values can be seen over cities of western Europe by ~20% and over eastern Europe by 30-50% (Fig. 5b). The tropospheric ozone data is not available over Europe (Fig. 5c). $PM_{2.5}$ concentrations show a reduction by ~33% throughout western Europe except Switzerland and by ~60-70% over eastern Europe, Sweden, Norway and Finland (Fig. 5d). However, a point to note is the miniscule absolute reductions and the base values making relative changes over countries such as Norway insignificant. We also see anomalous improvements in air quality over Switzerland are significant, however only over major cities such as Zurich, Lucern and Basel. The baseline in other regions, particularly the mountains is inherently low. However, looking at the absolute values from Google Earth Engine apps, it seems that the lockdown was highly effective as the $NO_2$ values can be seen to have dropped down to near zero.

### 3.6 North America

The region was earlier a major source of emissions, however with strict emission regulations in the last few decades there has been a marked reduction in rate of increment in air pollutants ($NO_x$, $PM_{2.5}$) over the high-income regions of North America (Canada, United States) (Naghavi et al. 2015). Results from past studies show that AOD and $PM_{2.5}$ agree in terms of interannual variability and both have decreased with time. Strongest changes have been noted over the eastern part with moderate changes over the central and western parts (Li et al. 2015). During the analysis period, we notice further reduction in $NO_2$ concentrations over North America which is visible in patches hovering over and around the urban areas. The East coast of the United States of America showed enhanced reductions in $NO_2$ relative to the west coast. Prominent reductions can be seen over New York, Atlanta, Charlotte, Detroit, Chicago, Denver, Los Angeles, San Jose, Portland, Calgary, Edmonton, Toronto, Montreal and Mexico City by ~30% (Fig. 6a). Other regions around these major centres of decreasing $NO_2$ show ~20% decrease in $NO_2$. Similar characteristics can be seen in the change in AOD and PM$_{2.5}$ concentrations (Fig. 6b,d). Tropospheric ozone data is absent over North America in our analysis (Fig. 6c). We particularly note a large increase in NO$_2$ in Western Canada, much more extended than in the highly populated Eastern side of the US. This is because the figure shows relative change, if however, we compare the absolute values, it is much less as compared to Eastern US. The state of Alberta and the city of Edmonton and its surroundings in particular have shown substantial improvements in air quality even in terms of absolute values.

### 3.7 South America

The region is dominated by the seasonal variability in air pollutant levels where biomass burning contributes maximum to the NO$_x$ variability (Castellanos et al. 2014). Concentrations of PM$_{2.5}$ and NOx in the urban parts are mostly influenced by the traffic being the main source (Krecl et al., 2018). We note reduction in NO$_2$ levels over large swathes of Brazil, Argentina, Chile, Peru, Columbia and Venezuela by ~20-30% (Fig. 7a). A 33% decrease in AOD can be observed over south-eastern Brazil, southern Argentina, Bolivia and Peru with the lowered values reaching 50% concentrated in and around Sao Paulo (Fig. 7b). We can see 15-20% decrease in tropospheric ozone concentrations over Brazil and Bolivia (Fig. 7c). Large parts of South America show PM$_{2.5}$ reductions upto ~60-70% with some regions such as the Roraima in Brazil and Santa Cruz, Chubut in Argentina showing complete removal of PM$_{2.5}$ (Fig. 7d). An important point to note is the improved air quality in South Argentina which is poorly populated but has considerable oil production. We note as also from (https://www.spglobal.com/platts/en/market-insights/latest-news/oil/062620-argentina-extends-tightens-lockdown-puts-fresh-damper-on-oil-demand-production) that Argentina went into an extended period of lockdown and it also consumes most of its oil. These might have been the reasons for improved air quality in those regions.

### 3.8 Southeast Asia

This region is also subjected to strong seasonal cycles in air pollutants due to changes in the atmospheric circulation patterns, however the increasing emissions have a significant contribution by the biomass burning from Peninsular Southeast Asia (Dong and Fu 2015; Wang et al., 2017).The finer mode $PM_{2.5}$ is mostly associated with human activities wherein the AODs peak during the biomass burning season (Su et al., 2010; Lalitaporn et al., 2013). It can be seen that during the lockdown period $NO_2$ concentrations reduced by ~30% over Hanoi, Malaysia, Singapore, Jakarta, Manila and large parts of Indonesia (Fig. 8a). Other regions in southeast Asia also show a decrease by ~20% of $NO_2$. Hanoi shows the maximum reduction in AOD by ~50% followed by Singapore (~30-40%) and marginal decrease in other parts of southeast Asia (Fig. 8b). Tropospheric ozone decreased by 20-30% throughout the region (Fig. 8c) and $PM_{2.5}$ decreased by ~60-70% uniformly over the area (Fig. 8d).

### 3.8 Megacities

More than half of the human population as of 2020 delves in urban areas. Past few decades have seen exponential growth in the number of these agglomerates and larger urban segments known as megacities. Megacities have large economies and are also sources of anthropogenic pollutants. In recent times, the health hazards posed by these pollutants in the megacities have become evident (Parrish et al., 2009). Urban areas contribute maximum to the rising emissions throughout the world, with more pronounced effects in the developing countries. With the rise in population and demands these regions are the hotspots of such air pollutants. In addition to the regional analysis, we also dig into the lockdown forced changes in the megacities around the world. We note a significant reduction in the pollutants considered in the study. The results are present in the supplementary Table 2. We observe a statistically significant fall (Fig. 9) in the $NO_2$, AOD and $PM_{2.5}$ concentrations (Table 2) with the reductions amounting to 19.74%, 7.38% and 49.9% as the mean with respect to the same period in 2019 as lockdowns in the respective megacities in 2019. Fig. 9 shows the violin plots of the various species wherein the bubbles represent the distribution of the global changes. We also note that the pandemic forced lockdown led to limit the air pollution levels close to or below the targeted levels as prescribed by the WHO and UNEP led air pollution monitoring network as part of the Global Environment Monitoring System (Mage et al., 1996). Lima, Port Louis and Mumbai are the top three cities with regards to reductions in $NO_2$ concentrations ~72%, ~56% and ~50% during 2020 lockdown relative to 2019 whereas Bogota, Guangzhou and Shenzen showed nominally enhanced $NO_2$ concentrations during lockdown. Jakarta, Kuala Lumpur and Bangkok are the top three cities with regards to 2020 lockdown reductions in $O_3$ concentrations to the tune of 16%, 12% and 10% relative to 2019, whereas Panama City, Mexico City and Manila show enhanced $O_3$ concentrations relative to 2019. Riyadh, Manama and Abu Dhabi show the largest reductions (75.6%, 73% and 72% respectively) in $PM_{2.5}$ concentrations during lockdown whereas Nassau is the only megacity showing slightly enhanced (~17%) $PM_{2.5}$ concentrations during lockdown. The megacities Shenzen, Sao Paulo and Luxembourg show maximum decrease (55%, 48% and 46% respectively) in AOD whereas Brussels, Panama City and Chongqing show an increase in AOD during the lockdown relative to 2019.

### 3.9    Google Earth Engine apps

A set of six Google Earth Engine apps have been developed to aid in enhanced visualization of the changes/improvements in air quality due to COVID-19 lockdowns. For best visualization of these apps, the use of Google Chrome browser is recommended. They can be accessed from the links below:

AOD: https://manmeet20singh15.users.earthengine.app/view/aodlockdown
$NO_2$: https://manmeet20singh15.users.earthengine.app/view/no2lockdown
Tropospheric ozone: https://manmeet20singh15.users.earthengine.app/view/tropospherico3lockdown

PM$_{2.5}$: https://manmeet20singh15.users.earthengine.app/view/pm25lockdown
Land surface temperature: https://manmeet20singh15.users.earthengine.app/view/lstlockdown
Surface winds: https://manmeet20singh15.users.earthengine.app/view/windlockdown

## 4. Conclusions

In this study we have used remote sensing datasets accessed via Google Earth Engine to assess and quantify the impact of COVID-19 enforced lockdowns on the air pollutants across the world and over megacities in particular. This work has been possible only because of Google Earth Engine and also to the open access SENTINEL-5P and MODIS data from the European Space Agency and NASA. As compared to the air pollutants concentration for the same period in the previous year, we find significant reductions over all regions in the parameters such as $NO_2$, AOD, tropospheric ozone and PM$_{2.5}$. However, there are some anomalies or relative variations as well as significant differences between $NO_2$ and PM$_{2.5}$ variations that could justify further work to understand the reasons behind the variations. The satellite data are not available for 2018 and previous years and hence the comparison has only been done relative to 2019. The comparison of the COVID-19 enforced lockdown period to post-COVID-19 period would test the statistical significance of our results which can be taken in a follow-up study. The main features of this study are (i) use of complete lockdown period as compared to other works with limited time duration, (ii) use of multi-pollutants datasets, (iii) use of spatially contiguous satellite datasets enabling better understanding and (iv) understanding spatial distribution of the changes in air pollutants due to lockdown. As seen from the analysis, the consistent variation over most big cities is a good indication of the relationship between pollution decrease and COVID-19.

Further research can use the wind and temperature datasets to analyse the relationship. This study gives equal importance to Africa, which is less covered by observational networks and hence the information coming out of the region is limited (Dinku et al., 2019). The state-of-the-art climate models have large uncertainties in simulating these air pollutants, and COVID-19 has provided a testbed for their validation. If we look at Angola, Namibia, South Africa, Australia and New Zealand, the decrease in NO2 concentration and PM2.5 is almost anti-correlated. Since both are only linked to human and particularly traffic reduction, further research needs to be carried out on this aspect. Moreover, in some areas away from megacities large variations or anomalous variations may not be significant if absolute values are low. This work can serve as a benchmark to assess the climate model simulations understanding the role of lockdown on air quality and hence can also be used to improve the climate model parameterizations.

## 5. Software Availability

The codes used in this study are available as an open source version from https://github.com/manmeet3591/gee_lockdown

**Fig. 1 COVID19 lockdown changes in atmospheric pollutants over Africa and Middle East**: Spatial maps of percentage change in concentrations of (**a**) $NO_2$ (**b**) AOD (**c**) Tropospheric $O_3$ and (**d**) $PM_{2.5}$ for the 2020 COVID19 enforced lockdown period relative to the same period in 2019. The change in concentration is represented as (period in 2019 corresponding to the lockdown in 2020 - 2020 COVID19 enforced lockdown) expressed in percentage relative to the period in 2019 corresponding to the lockdown in 2020. The data used is from the TROPOMI instrument onboard Sentinel-5P satellite, MODIS and Sentinel-2 MultiSpectral Instrument.

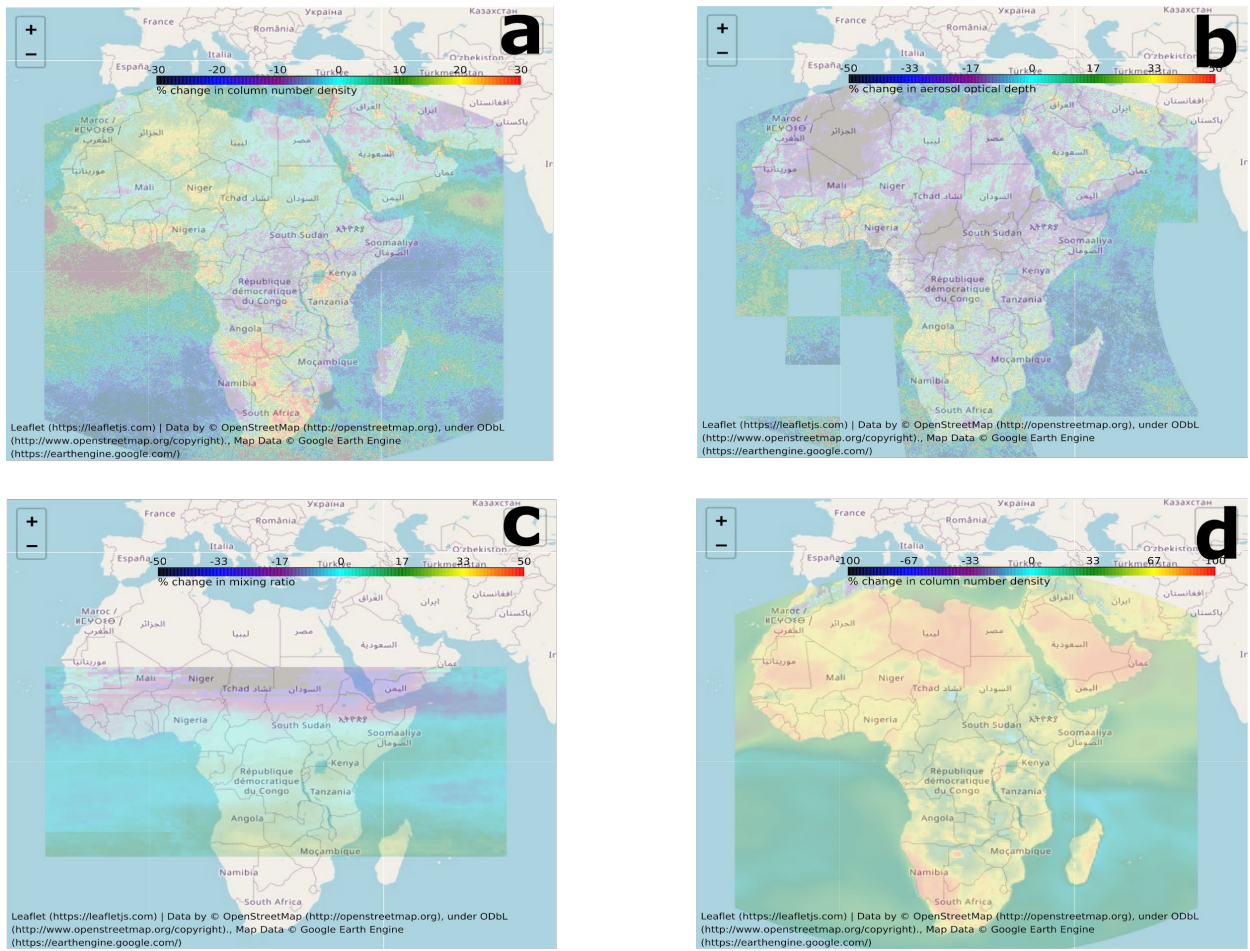

**Fig. 2 COVID19 lockdown changes in atmospheric pollutants over Australia and New Zealand**: Spatial maps of percentage change in concentrations of (**a**) $NO_2$ (**b**) AOD (**c**) Tropospheric $O_3$ and (**d**) $PM_{2.5}$ for the 2020 COVID19 enforced lockdown period relative to the same period in 2019. The change in concentration is represented as (period in 2019 corresponding to the lockdown in 2020 - 2020 COVID19 enforced lockdown) expressed in percentage relative to the period in 2019 corresponding to the lockdown in 2020. The data used is from the TROPOMI instrument onboard Sentinel-5P satellite, MODIS and Sentinel-2 MultiSpectral Instrument.

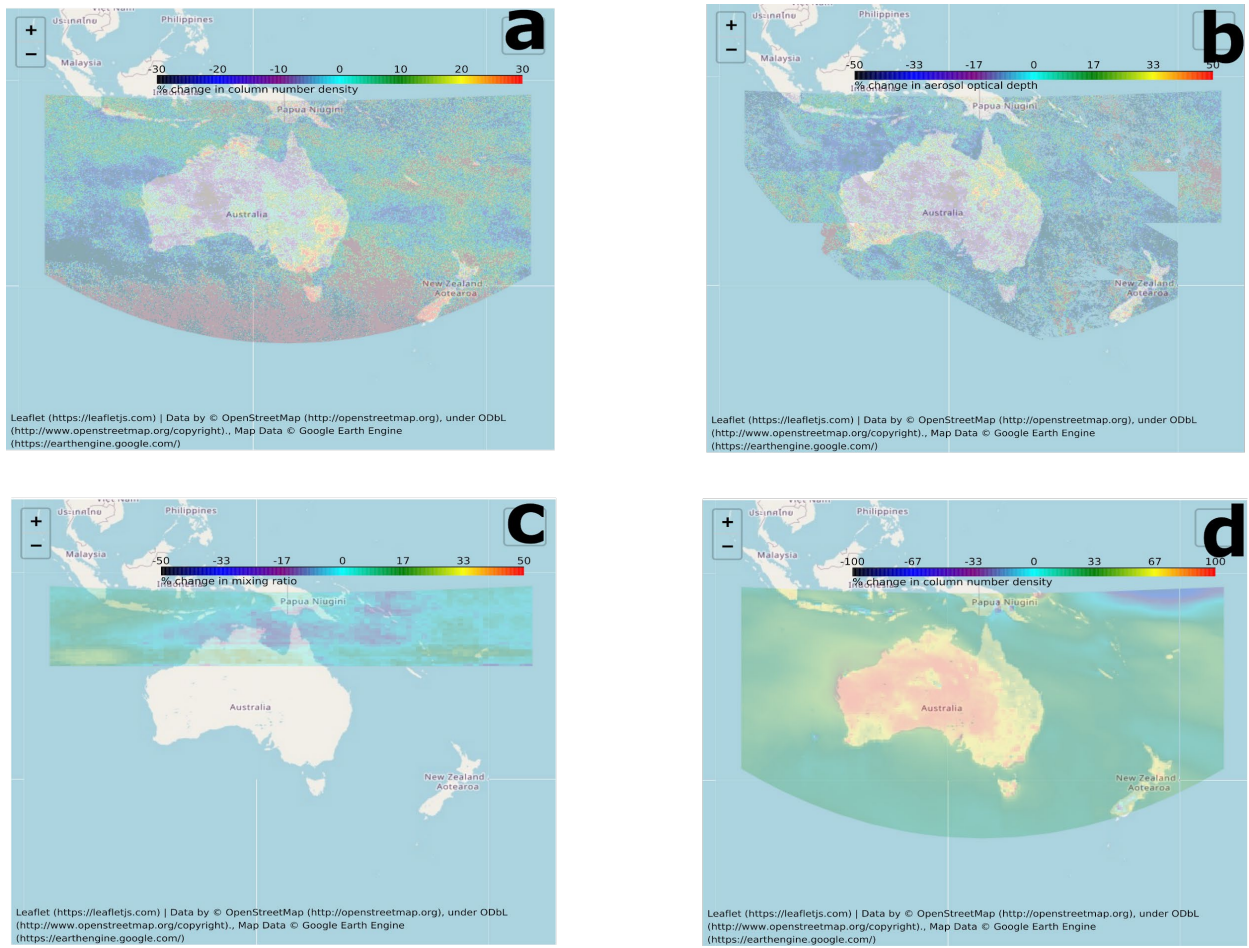

**Fig. 3 COVID19 lockdown changes in atmospheric pollutants over East Asia**: Spatial maps of percentage change in concentrations of (**a**) NO$_2$ (**b**) AOD (**c**) Tropospheric O$_3$ and (**d**) PM$_{2.5}$ for the 2020 COVID19 enforced lockdown period relative to the same period in 2019. The change in concentration is represented as (period in 2019 corresponding to the lockdown in 2020 - 2020 COVID19 enforced lockdown) expressed in percentage relative to the period in 2019 corresponding to the lockdown in 2020. The data used is from the TROPOMI instrument onboard Sentinel-5P satellite, MODIS and Sentinel-2 MultiSpectral Instrument. Note that the color scale for relative change is from -50% to +50%.

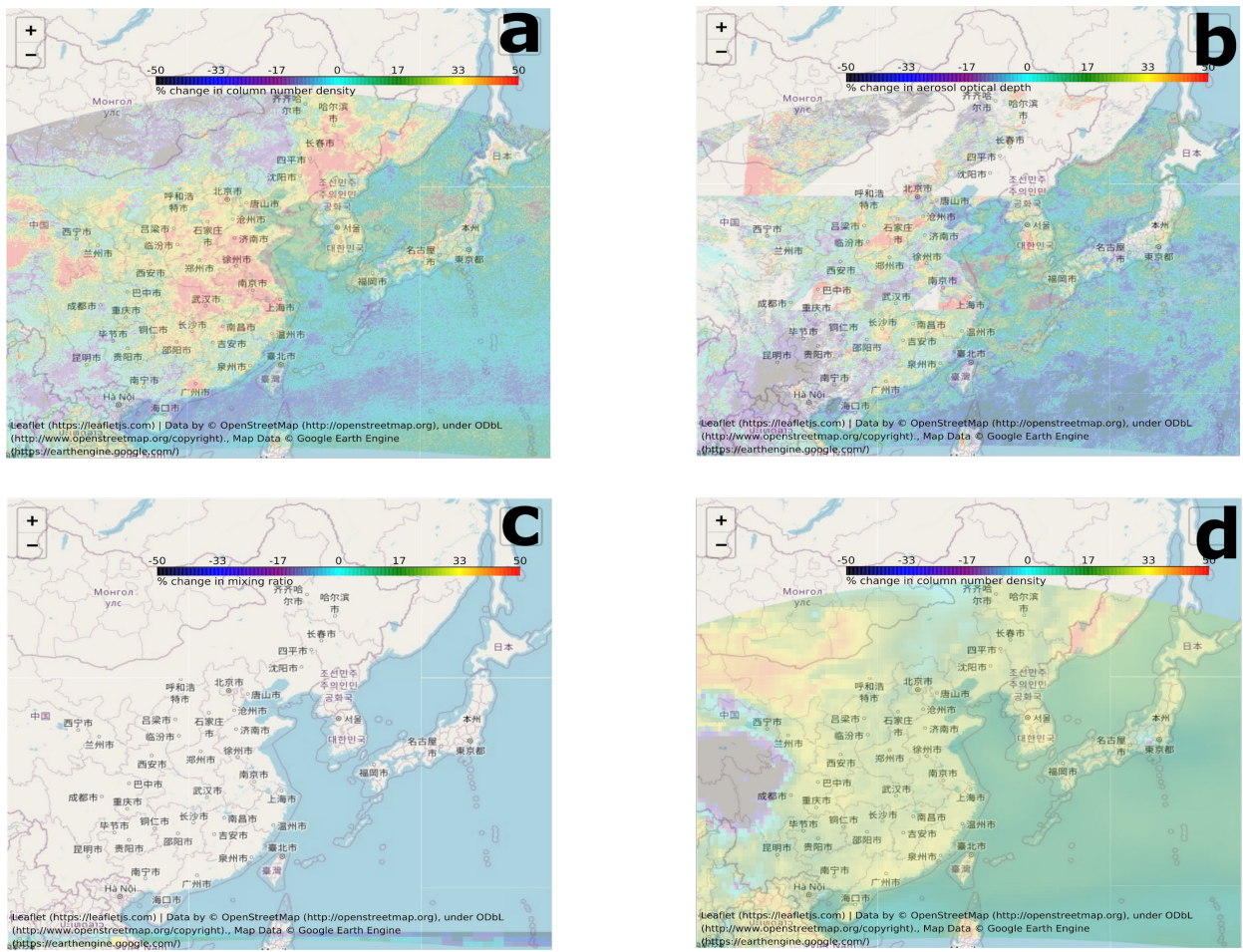

**Fig. 4 COVID19 lockdown changes in atmospheric pollutants over South Asia**: Spatial maps of percentage change in concentrations of (**a**) $NO_2$ (**b**) AOD (**c**) Tropospheric $O_3$ and (**d**) $PM_{2.5}$ for the 2020 COVID19 enforced lockdown period relative to the same period in 2019. The change in concentration is represented as (period in 2019 corresponding to the lockdown in 2020 - 2020 COVID19 enforced lockdown) expressed in percentage relative to the period in 2019 corresponding to the lockdown in 2020. The data used is from the TROPOMI instrument onboard Sentinel-5P satellite, MODIS and Sentinel-2 MultiSpectral Instrument.

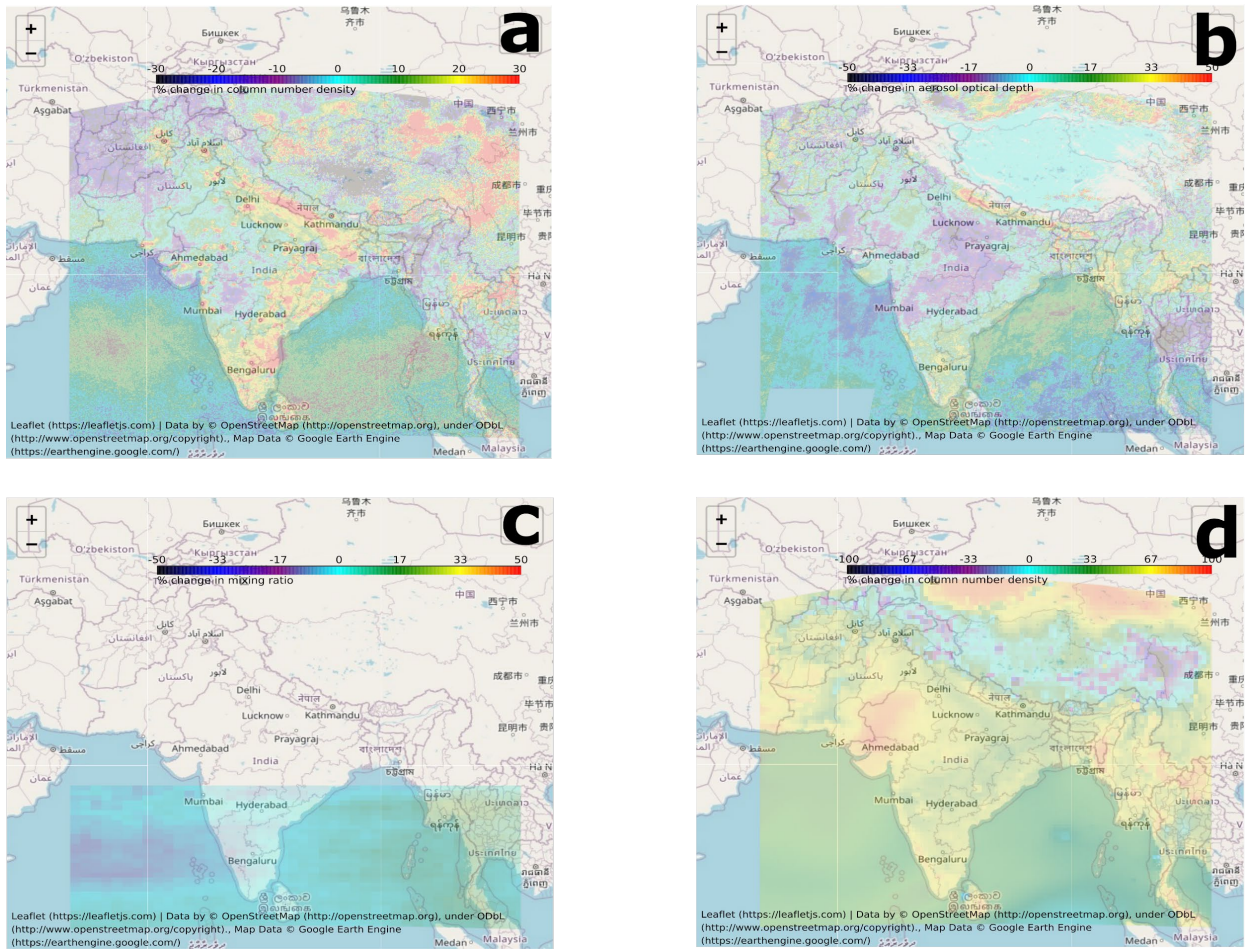

**Fig. 5 COVID19 lockdown changes in atmospheric pollutants over Europe**: Spatial maps of percentage change in concentrations of (**a**) NO$_2$ (**b**) AOD (**c**) Tropospheric O$_3$ and (**d**) PM$_{2.5}$ for the 2020 COVID19 enforced lockdown period relative to the same period in 2019. The change in concentration is represented as (period in 2019 corresponding to the lockdown in 2020 - 2020 COVID19 enforced lockdown) expressed in percentage relative to the period in 2019 corresponding to the lockdown in 2020. The data used is from the TROPOMI instrument onboard Sentinel-5P satellite, MODIS and Sentinel-2 MultiSpectral Instrument.

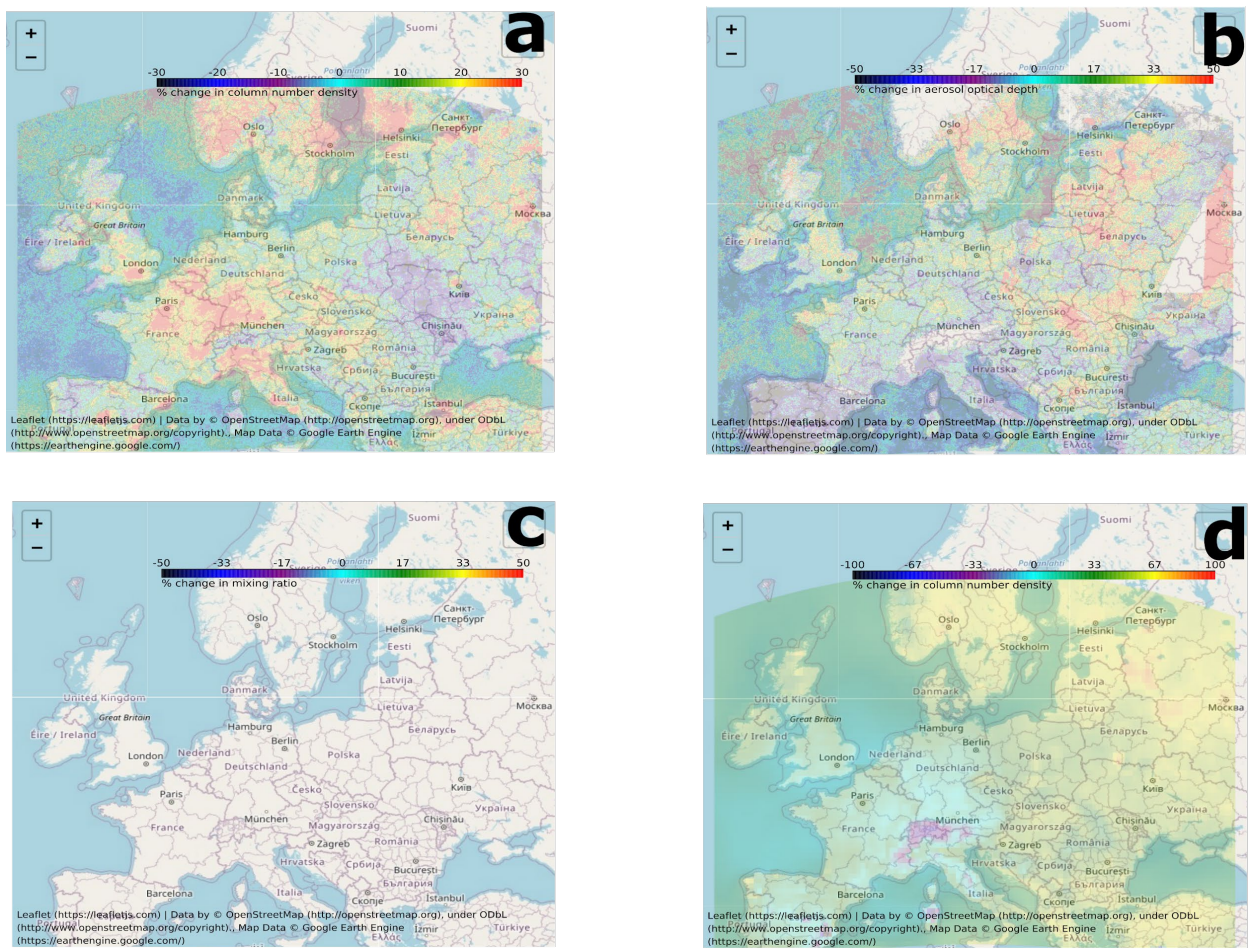

**Fig. 6 COVID19 lockdown changes in atmospheric pollutants over North America**: Spatial maps of percentage change in concentrations of (**a**) NO$_2$ (**b**) AOD (**c**) Tropospheric O$_3$ and (**d**) PM$_{2.5}$ for the 2020 COVID19 enforced lockdown period relative to the same period in 2019. The change in concentration is represented as (period in 2019 corresponding to the lockdown in 2020 - 2020 COVID19 enforced lockdown) expressed in percentage relative to the period in 2019 corresponding to the lockdown in 2020. The data used is from the TROPOMI instrument onboard Sentinel-5P satellite, MODIS and Sentinel-2 MultiSpectral Instrument.

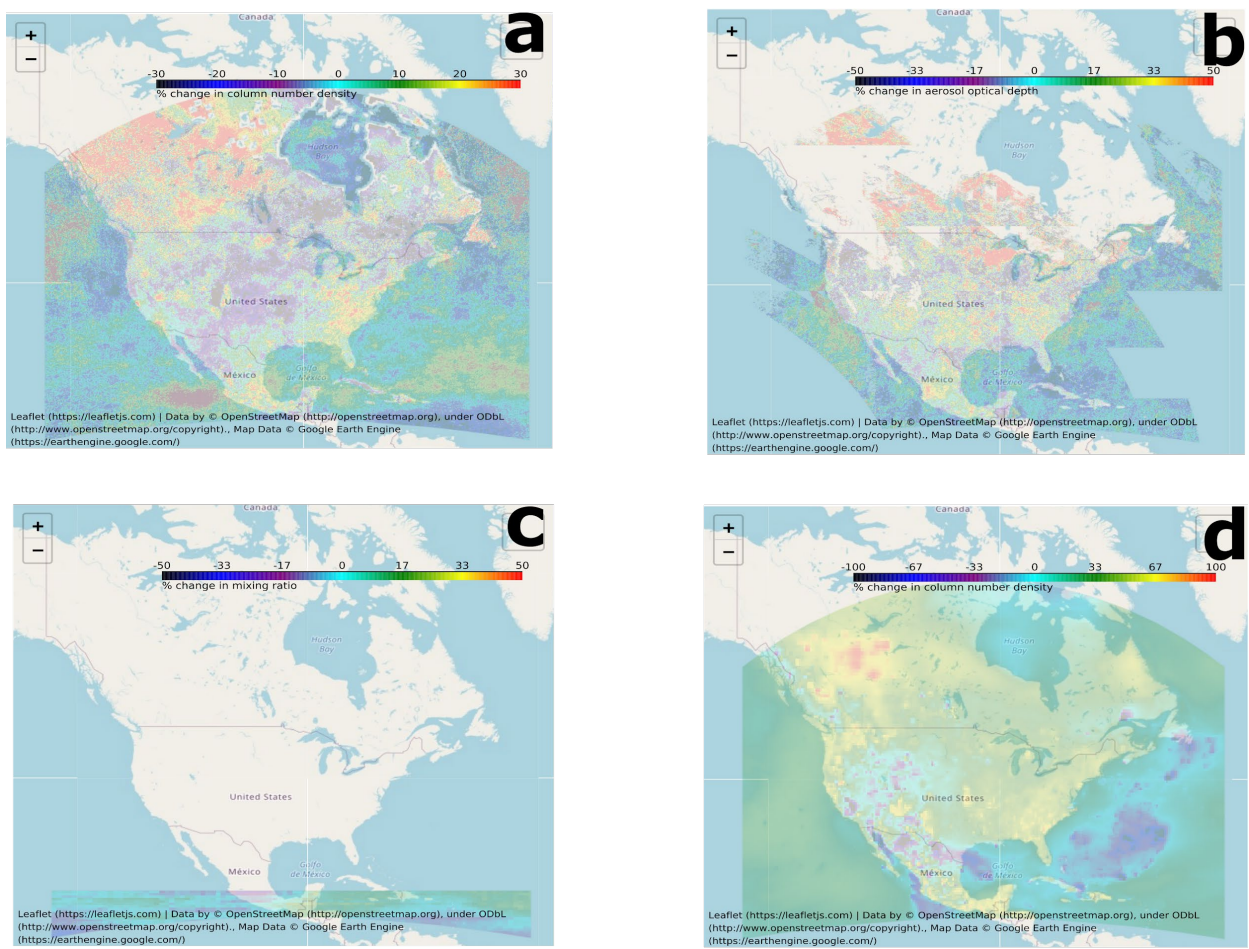

**Fig. 7 COVID19 lockdown changes in atmospheric pollutants over South America**: Spatial maps of percentage change in concentrations of (**a**) NO$_2$ (**b**) AOD (**c**) Tropospheric O$_3$ and (**d**) PM$_{2.5}$ for the 2020 COVID19 enforced lockdown period relative to the same period in 2019. The change in concentration is represented as (period in 2019 corresponding to the lockdown in 2020 - 2020 COVID19 enforced lockdown) expressed in percentage relative to the period in 2019 corresponding to the lockdown in 2020. The data used is from the TROPOMI instrument onboard Sentinel-5P satellite, MODIS and Sentinel-2 MultiSpectral Instrument.

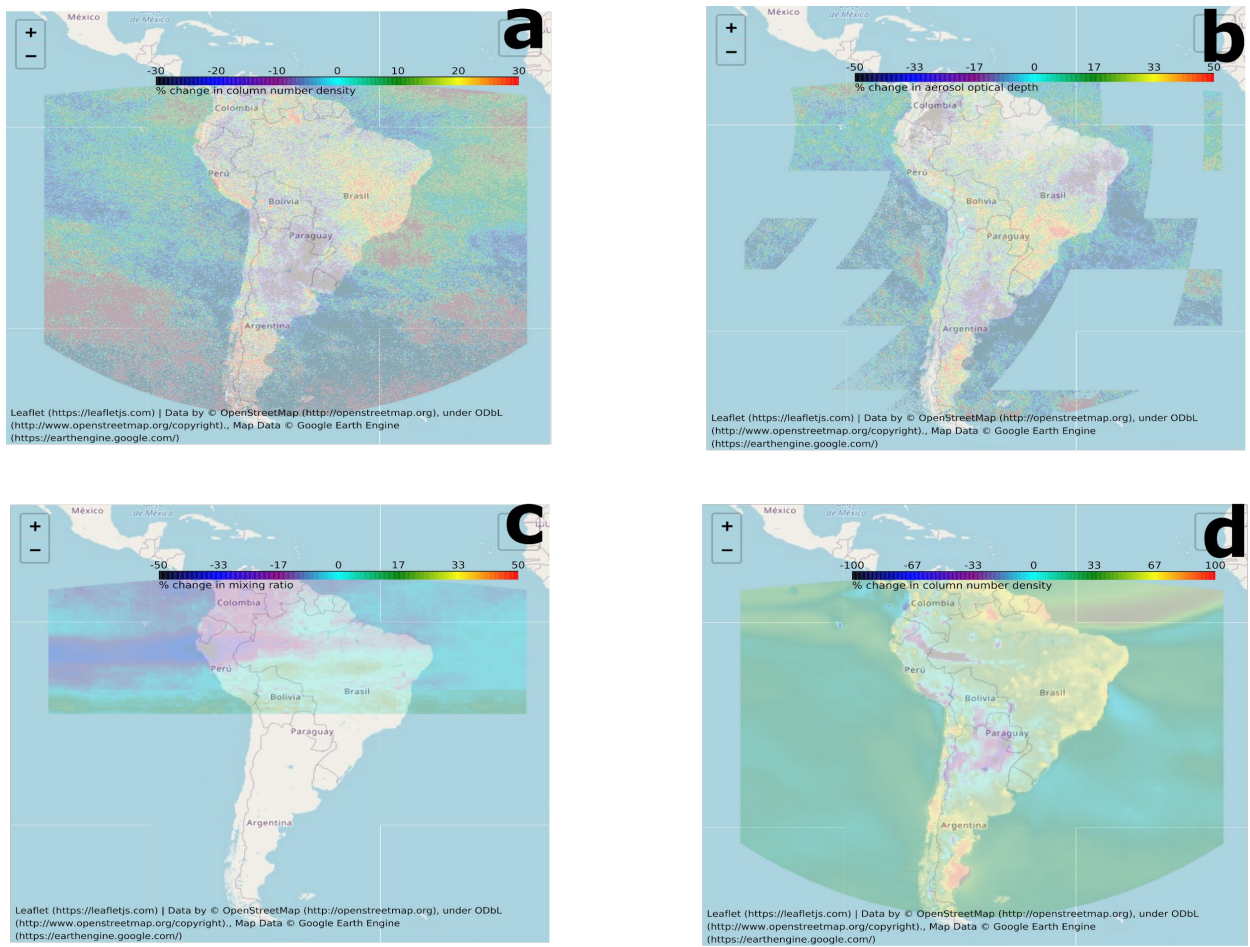

**Fig. 8 COVID19 lockdown changes in atmospheric pollutants over Southeast Asia**: Spatial maps of percentage change in concentrations of (**a**) NO$_2$ (**b**) AOD (**c**) Tropospheric O$_3$ and (**d**) PM$_{2.5}$ for the 2020 COVID19 enforced lockdown period relative to the same period in 2019. The change in concentration is represented as (period in 2019 corresponding to the lockdown in 2020 - 2020 COVID19 enforced lockdown) expressed in percentage relative to the period in 2019 corresponding to the lockdown in 2020. The data used is from the TROPOMI instrument onboard Sentinel-5P satellite, MODIS and Sentinel-2 MultiSpectral Instrument.

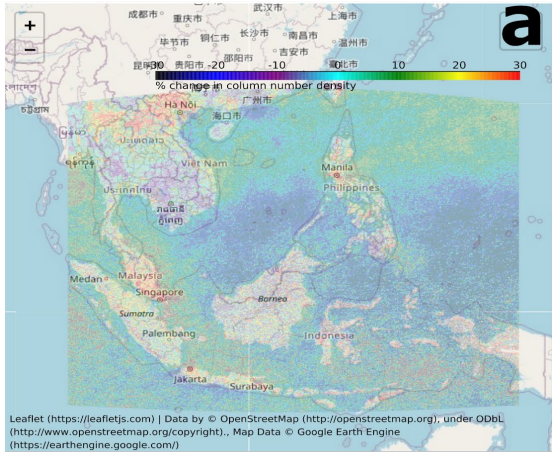 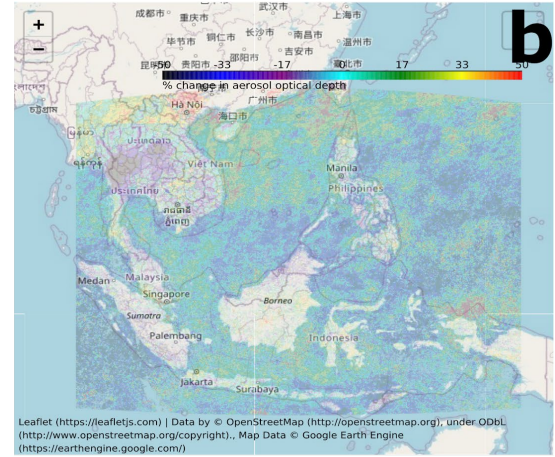
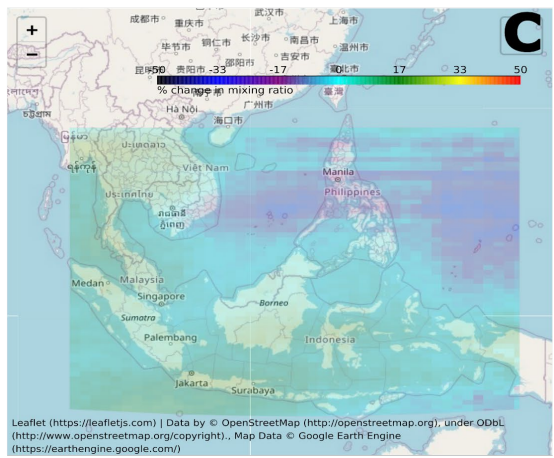 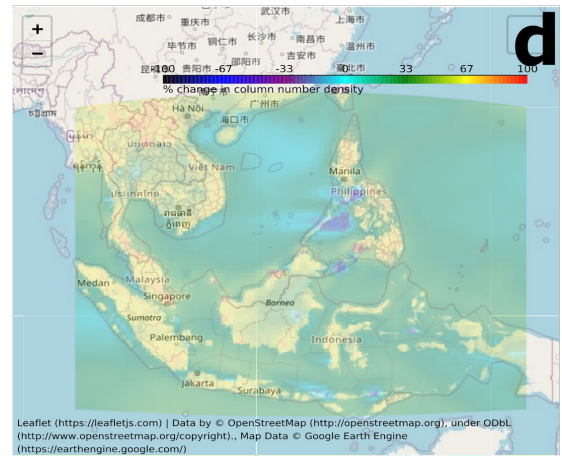

**Fig. 9** Distributions of percentage change in NO$_2$ concentrations, AOD, Tropospheric ozone and PM$_{2.5}$ of global megacities for the year 2019 relative to 2020

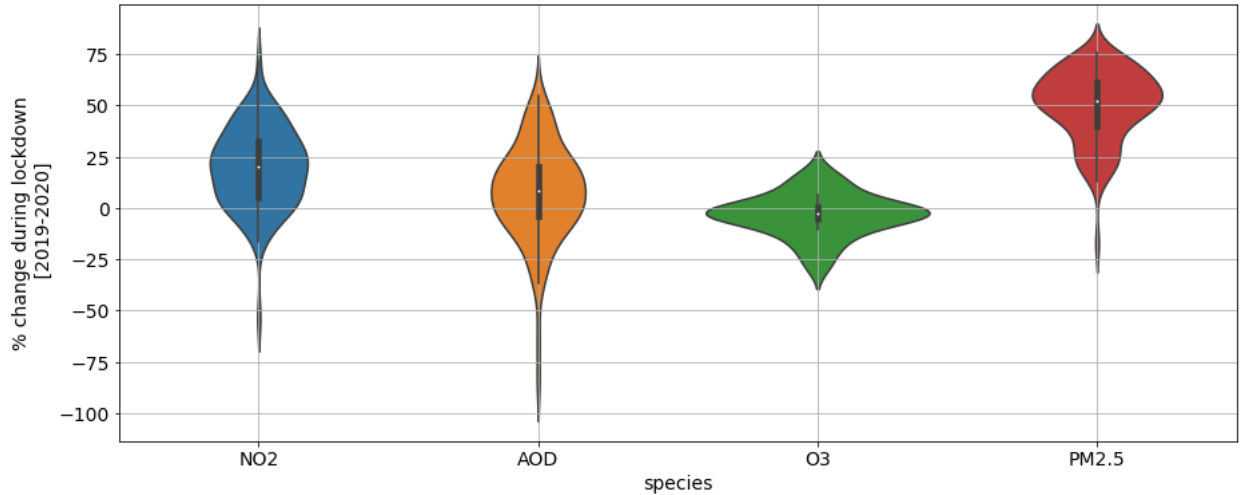

**Table 1.** Timelines of COVID19 enforced lockdowns in various regions of the world obtained from news articles and national reports

| Lockdown period | Region |
| --- | --- |
| 23 March 2020 to 31 May 2020 | South Asia (India and surrounding regions) |
| 23 March 2020 to 31 May 2020 | South East Asia (Thailand, Malaysia, Singapore and surrounding regions) |
| 24 January 2020 to 25 March 2020 | East Asia (China, Japan, Korea and neighbourhoods) |
| 23 March 2020 to 31 May 2020 | Australia and New Zealand |
| 13 March 2020 to 31 May 2020 | Europe |
| 23 March 2020 to 31 May 2020 | North America |
| 13 March 2020 to 31 May 2020 | South America |

| 30 March 2020 to 31 May 2020 | Africa and Middle East |

**Table 2.** Percentage change in $NO_2$ concentrations, AOD, Tropospheric ozone and PM2.5 of global megacities for the year 2019 relative to 2020. The first column shows species, the second column represents the p-value after performing 1 sample t test to test the statistical significance of change in air quality parameters. The null hypothesis is that the mean change in percentage is 0. The third column is the mean percentage change for the year 2019 relative to 2020.

| Species | p-value | Mean (% change) |
| --- | --- | --- |
| $NO_2$ | 0.0 | 19.74 |
| AOD | 0.004 | 7.38 |
| Tropospheric $O_3$ | 0.15 | -3.23 |
| $PM_{2.5}$ | 0.0 | 49.9 |

observations of the atmospheric composition for climate, air quality and ozone layer applications. Remote Sensing of Environment 120, 70–83. doi:10.1016/j.rse.2011.09.027

Venter, Z.S., Aunan, K., Chowdhury, S., Lelieveld, J., 2020. COVID-19 lockdowns cause global air pollution declines with implications for public health risk. doi:10.1101/2020.04.10.20060673

Wang, J., Xing, J., Mathur, R., Pleim, J.E., Wang, S., Hogrefe, C., Gan, C.-M., Wong, D.C., Hao, J., 2017. Historical Trends in $PM_{2.5}$-Related Premature Mortality during 1990–2010 across the Northern Hemisphere. Environmental Health Perspectives 125, 400–408. doi:10.1289/ehp298

World Urbanization Prospects 2018: Highlights, 2019. doi:10.18356/6255ead2-en

World Health Organization, 2013. Review of evidence on health aspects of air pollution – REVIHAAP Project 309.

World Health Organization, 2020. Weekly epidemiological update - 15 December 2020. https://www.who.int/publications/m/item/weekly-epidemiological-update---15-december-2020.

Xu, K., Cui, K., Young, L.-H., Wang, Y.-F., Hsieh, Y.-K., Wan, S., Zhang, J., 2020. Air Quality Index, Indicatory Air Pollutants and Impact of COVID-19 Event on the Air Quality near Central China. Aerosol and Air Quality Research 20, 1204–1221. doi:10.4209/aaqr.2020.04.0139
Preprint accepted in Remote Sensing Applications: Society and Environment          38